\documentclass[prd,tightenlines,nofootinbib,amsfonts,amssymb,amsmath,color,11pt]{article}
\usepackage[english]{babel}
\usepackage[affil-it]{authblk}
\usepackage[T1]{fontenc}
\usepackage[latin9]{inputenc}
\usepackage{amsmath}
\usepackage{babel}
\usepackage{atbegshi}
\AtBeginDocument{\AtBeginShipoutNext{\AtBeginShipoutDiscard}}
\usepackage{babel}
\usepackage{amssymb}
\usepackage{hyperref}
\usepackage{geometry}
\usepackage{color}
\usepackage{graphicx}
\usepackage{subfig}
\usepackage[font=small,labelfont=bf]{caption}
\usepackage{hyperref}
\hypersetup{colorlinks=true, citecolor=blue, urlcolor=blue, linkcolor=blue}
\usepackage{verbatim}
\global\long\def\imsize{0.83\columnwidth}
 \global\long\def\halfsize{0.45\columnwidth}

\begin{document}
\twocolumn
\title{\bf A note on splitting solutions in $4+1$ dimensional quadratic gravity}
\author[1]{Daniel M\"uller} 
\thanks{dmuller@unb.br}
\author[2]{Alexey Toporensky \thanks{atoporensky@gmail.com}}
\affil[1]{Instituto de F\'{i}sica, Universidade de Bras\'{i}lia, Caixa Postal
04455, 70919-970 Bras\'{i}lia, Brazil}
\affil[2]{Sternberg Astronomical Institute, Moscow University, Moscow 119991, Russian Federation}
\date{\today}
\maketitle
\abstract{{\small
In the present paper we consider anisotropic cosmological vacuum solutions in (4+1) dimensional general quadratic gravity. In particular, we present a solution with 3 equal and 1 different Hubble parameters, and study its stability. We show that for a certain range of coupling constants this solution is stable. This means that initially totally anisotropic 4-dim Universe
can evolve naturally to a product of 3-dim isotropic subspace and 1-dim space. By numerical integration of equations of motion we construct bassin of attraction of this solution which covers part of the initial conditions space with non-zero measure.
}}

\section{Introduction}
The problem of dynamical compactification is one of the most important in multi-dimensional cosmology. Indeed, in order to describe our world a multi-dimensional model should contain an almost isotopic 3-dim sub space (our observed universe) and a sub space of ``inner'' dimensions with almost constant volume. The reason for the least requirement is that the volume of the ``inner'' space can enter in the expressions for the effective values of fundamental constant such as the Newton constant which is known to be a constant up to high precision. 

To study this problem the metric is often chosen to be a product of two isotropic subspaces --
3-dimensional ``our'' world and an additional ``inner'' subspace. Since in such a set-up there are only two different scale factors, the resulting cosmological dynamics can be studied analytically in most cases \cite{Tosa, Gud, Randall, Kim}.
It is already known that the requirement of the ``inner'' sub space stabilization can be rather easily fulfilled in modify gravity theories. A spatial curvature of the ``inner'' space usually stabilizes it, though other types of solution including singular solutions are possible. Nevertheless, stabilization does not require any serious fine-tinning of initial conditions. This has been shown for Gauss-Bonnet gravity and Weyl gravity \cite{Giacomini1, Giacomini2, Shapiro}.

However, such a choice of the metric from the beginning means that we already separate 3-dim world ``by hand'' in the study, and it would be desirable to get this separation dynamically starting from some general anisotropic metric. Since there are too many possibilities to write down a multi-dimensional metric with spatial curvature, without any canonical pattern valid for any dimensions, the simplified problem would be to start from a general anisotropic flat metric (i.e., multi-dimensional generalization of Bianchi I metric ), and search for an outcome being a sum of isotopic 3-dim and a contracting ``inner'' spaces.
After such splitting into two isotopic sub spaces occurs, further evolution can be traced using the former set up with only two different scale factor, and taking a spatial curvature into account at this stage is not a problem. Physically, spatial curvature evidently becomes more important at contraction, allowing us to consider it only at this second stage of evolution after the splitting takes place. 
After such simplification, it appears that multi-dimensional dynamics in Gauss- Bonnet cosmology satisfy our needs. Non singular anisotropic trajectories of a flat background usually end up by splitting into two isotopic de Sitter spaces \cite{Chirkov1, Chirkov2}. If the Hubble parameter of the second de Sitter space is negative, further evolution (when curvature is already put into account) leads to stabilization of the ``inner'' space \cite{Pavl1, Pavl2}. By choosing coupling constants it is possible to set the Hubble parameter of ``our'' world close to zero in natural units also \cite{Friedmann}.

The goal of the present paper is to consider possibilities of similar scenarios in  general quadratic gravity. 

Quadratic gravity naturally arises as vacuum polarization effect of quantized matter fields in classical background gravitational fields, 
\cite{UdW} see also \cite{dw} and \cite{Grib} for instance. As is well known, when itense fields are present, the quantized Dirac field polarizes vacuum \cite{Schwinger} and Euler-Heisenberg electrodynamics is recovered \cite{E-H}. That is the reason that this technique is known as the Schwinger-de Witt method, and we believe that in the very early Universe, the gravitational field is sufficiently intense that quadratic gravity may be a good candidate, at least in an effective sense.

Formally, solutions in the form of sum of two de Sitter sub spaces exist, and it is rather simple to find them analytically. It is stability which is a problem. General quadratic gravity is a fourth order theory in contrast to   Gauss-Bonnet gravity which is  a second order theory, so the number of degrees of freedom is bigger, and extra degrees of freedom could be responsible for turning such a solution to be unstable. We remind a reader that in Gauss-Bonnet gravity any solution with constant Hubble parameters is stable if the overall expansion rate is positive \cite{Ivashchuk}. We do not expect that such a "miracle" appears in more general theories.  Since the equations of motion in a general quadratic gravity are rather cumbersome, we restrict ourselves here by the simplest case of 4+1 dimension metrics.
\section{The theory}
The line element is the appropriate one for 4D-flat space 
\begin{align}
    ds^2=-dt^2+e^{2a_1}dx^2+e^{2a_2}dy^2+e^{2a_3}dz^2+e^{2b_1}du^2
\end{align}
where we immediately recognize the Hubble parameters,
\begin{align}
    H_1=\dot{a}_1 && H_2=\dot{a}_2 && H_3=\dot{a}_3 && h_1=\dot{b}_1
\end{align}
As explained in the Introduction the theory supposed 
\begin{align}
    L=\sqrt{-g}\left(\frac{R-2\Lambda}{8\pi G}+\alpha R_{ab}R^{ab}+\beta R^2 +\gamma R^{abcd}R_{abcd} \right)
\end{align}
is the most general in $4+1D$, as the Gauss-Bonnet invariant, $R_{abcd}R^{abcd}-4R_{ab}R^{ab}+R^2$, is not a surface term in 5D.   
 
 The field equations are obtained with respect to the $a_{c}$ coefficients
for 
\begin{align*}
 & \left[+\frac{d^{2}}{dt^{2}}\left(\frac{\partial L}{\partial\ddot{a}_{c}}\right)-\frac{d}{dt}\left(\frac{\partial L}{\partial\dot{a}_{c}}\right)+\frac{\partial L}{\partial a_{c}}\right]_{a_{0}=0}=0,
\end{align*}
which give respectively for $a_{c}$ for $c=0$, and $c=i=1,2,3,4$
\begin{align*}
 & E_{0}=\left[+\frac{d^{2}}{dt^{2}}\left(\frac{\partial L}{\partial\ddot{a}_{0}}\right)-\frac{d}{dt}\left(\frac{\partial L}{\partial\dot{a}_{0}}\right)+\frac{\partial L}{\partial a_{0}}\right]_{a_{0}=0}=0\\
 & E_{i}=\left[+\frac{d^{2}}{dt^{2}}\left(\frac{\partial L}{\partial\ddot{a}_{i}}\right)-\frac{d}{dt}\left(\frac{\partial L}{\partial\dot{a}_{i}}\right)+\frac{\partial L}{\partial a_{i}}\right]_{a_{0}=0}=0.
\end{align*}
which are written in the Appendix. As always, the $0$ equation is a constraint and is used as a numerical check.  
Metric variations 
\[
\frac{\delta}{\delta g_{ij}}=\frac{\delta a_{c}}{\delta g_{ij}}\frac{\delta}{\delta a_{c}}=\frac{1}{2g_{ik}}\delta_{kj}\frac{\delta}{\delta a_{j}}=\delta_{ij}\frac{e^{-2a_{j}}}{2}\frac{\delta}{\delta a_{j}}
\]
 for $i,j,k=1,2,3,4$ and 
\[
\frac{\delta}{\delta g_{00}}=\frac{1}{2g_{00}}\frac{\delta}{\delta a_{0}}=-\frac{e^{-2a_{0}}}{2}\frac{\delta}{\delta a_{0}},
\]
while $g_{i0}=0$ is zero. In this fashion, variations in $a_{i}$
\begin{align*}
 & E_{i}=\frac{\delta}{\delta a_{i}}L &  & E_{0}=\frac{\delta}{\delta a_{0}}L\\
 & T^{ij}=\delta_{ij}\frac{e^{-2a_{j}}}{2}E_{j} &  & T^{00}=\left[-\frac{e^{-2a_{0}}}{2}E_{0}\right]_{a_{0}=0},
\end{align*}
where $T_{ab}$ is the energy momentum tensor and in this case $T_{0i}=0.$
Now, as in any other metric theory of gravity, the energy momentum
tensor is covariantly conserved such that $\nabla_{a}T^{ab}=0$ for
$b=0$
\begin{align*}
\nabla_{a}T^{a0}=-\frac{1}{2}\dot{E}_{0}+\Gamma_{a0}^{a}T^{00}+\Gamma_{ac}^{0}T^{ac}=\\
-\frac{1}{2}\dot{E}_{0}+\left(\dot{a}_{1}+\dot{a}_{2}+\dot{a}_{3}+\dot{a}_{4}\right)\frac{E_{0}}{2}\\
+\frac{1}{2}\left(\dot{a}_{1}E_{1}+\dot{a}_{2}E_{2}+\dot{a}_{3}E_{3}+\dot{a}_{4}E_{4}\right)=0,
\end{align*}
remind that from the line element the non zero Christoffel symbols
are $\Gamma_{0i}^{i}=\dot{a}_{i}$ and $\Gamma_{ij}^{0}=\delta_{ij}e^{2a_{i}}\dot{a}_{i}$.
From this last equation it is possible to see that once the equations
of motion are satisfied, and initially the constraint $E_{0}=0$, the
constraint is dynamically maintained.

We stress again, since explicit expressions for $E_i$ and $E_0$ are rather cumbersome, we present them in Appendix.

\section{Splitting solutions and their stability}
This gravity theory allows for a (3+1) splitting solutions with three equal and one different constant Hubble parameters,
and all higher derivatives are zero. A natural way to present a solution would be to fix coupling constants and obtain expressions for Hubble parameters satisfying
the equations of motion. However, less combursome expressions arise when we fix Hubble parameters as well as two coupling constants $\alpha$ and $\gamma$ and express
$\Lambda$ and $\beta$ through them:
\begin{align}
 & \Lambda=-\left[\left(32\pi\alpha GH^{2}+1/2\right)h^{2}\right.\nonumber\\
 &\left.+\left(96\pi\alpha GH^{3}+3H/2\right)h+3H^{2}\right]\nonumber\\
 & \beta=\frac{32\pi(\alpha+\gamma)h^{2}+96\pi(\gamma-\alpha)Hh}{16\pi h^{2}+48\pi H^{2}}\nonumber\\
 &+\frac{32\pi(\alpha+6\gamma)H^{2}+1/(2G)}{16\pi h^{2}+48\pi H^{2}},\label{sol.de.sitter}
\end{align}
 where $H_{1}=H_{2}=H_{3}=H$ and $h_{1}=h.$ 
 
\begin{figure}
\begin{center}
\begin{tabular}{c c}
 \resizebox{\halfsize}{!}{\input{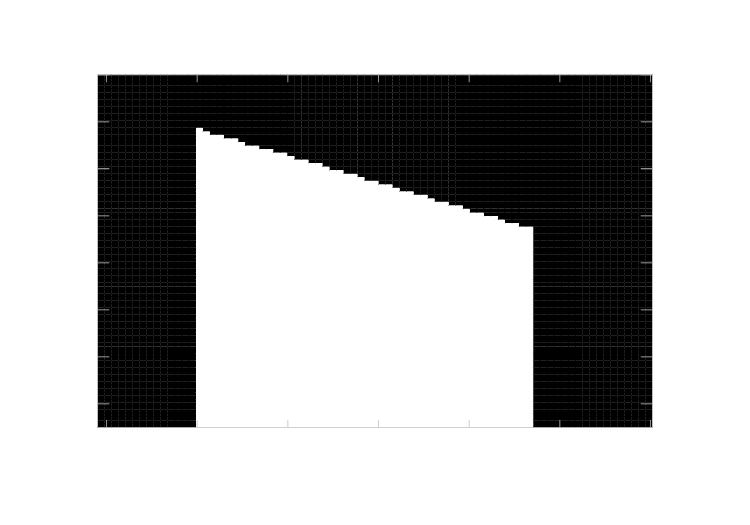}}   & 
 \resizebox{\halfsize}{!}{\input{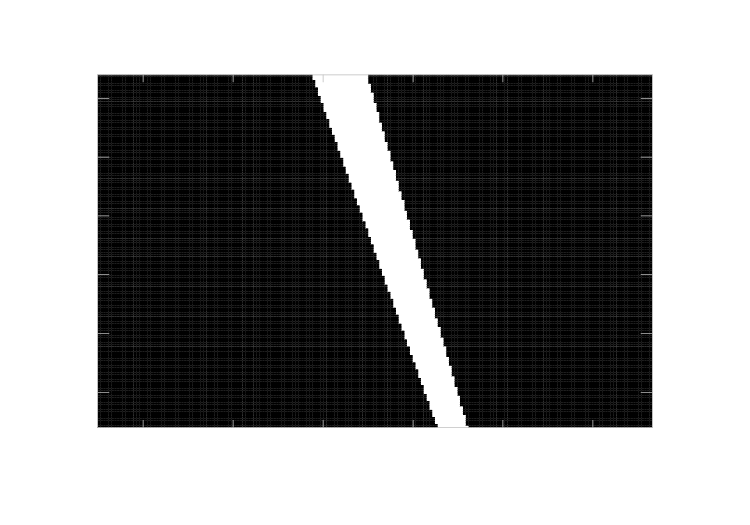}}  \\
       a) & b) 
   \end{tabular}
    \end{center}
    \caption{Stable region in parameter space. Panel a) Plane $\alpha\times \gamma$. For each white point all the real parts of the eigenvalues are negative, and each black point the real parts can be zero or positive. Zone of negative real part of all the eigenvalues is the ``good'' region which is marked by white. For this plot we chose $H=0.5201$ and $h=-0.2407$, which for each pair of values $\alpha,\,\gamma$ sets the value for $\beta$ and $\Lambda$ \eqref{sol.de.sitter}. Panel b) Now we fix $\alpha=0.02$ and $\gamma=-16$, which is a white point from panel a), and now a grid of points in $h,\,H$ which according to \eqref{sol.de.sitter} fixes the values for $\beta$ and $\Lambda$. \label{fig5}}
    \end{figure}

 Then we choose a negative
$h$ and a positive $H$ and look for eigenvalues of the linearized
system near this solution  for different values  of $\alpha$ and $\gamma$.
The equations of motion are of the 3-d order, so that we have in total 12 eigenvalues. If 
the real part of all of them are negative, the solution is known to be locally stable. A particular example is shown in Figure \ref{fig5}.
We evaluate eigenvalues for fixed $H=0.5201$ and $h=-0.2407$, which for each pair of values $\alpha,\,\gamma$ sets the value for $\beta$ and $\Lambda$ \eqref{sol.de.sitter}, shown in Figure \ref{fig5}a). While in Figure \ref{fig5}b) $\alpha$ and  $\gamma$  are set to $\alpha=0.02$ and $\gamma=-16$, which is a white point from panel a), and now a grid of points in $h,\,H$ which according to \eqref{sol.de.sitter} fixes the values for $\beta$ and $\Lambda$. The white region there corresponds to a situation when 
all 12 real parts of eigenvalues are negative. 
    Figure \ref{fig5} indicates that there exists a region on a plane of the coupling constants $\alpha$
    and $\gamma$ such that the (3+1) splitting solution is locally stable.

    For understanding the global stability we pick up a particular point in white region of Figure \ref{fig5}
    and trace numerically the fate of trajectories starting from different initial $H_i$. We consider that there are no relations between initial Hubble parameters, so we can vary them freely. All initial derivatives of $H_i$ are set to zero, except for $\ddot{h}_1$ which is calculated from the constraint equation. As the initial conditions space in such a setting is 4-dimensional, we present here two examples of 2-dim 
    section. 
    We see that there are two possible outcomes of the cosmological evolution - either splitting (black) or singular solution (white). Initial conditions leading to the splitting
    is seen to form rather nontrivial zone in the initial conditions space with clearly non-zero measure.
    
 \begin{figure}[htpb]
 \begin{center}
\begin{tabular}{c c}
 \resizebox{\halfsize}{!}{\input{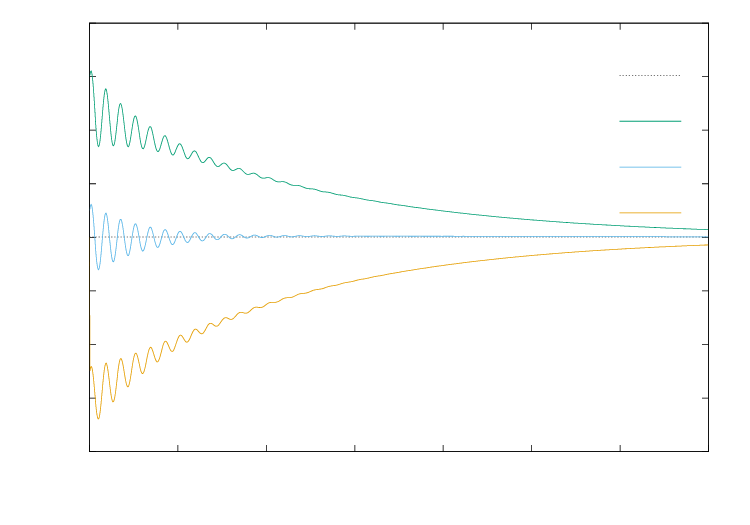}}   & 
 \resizebox{\halfsize}{!}{\input{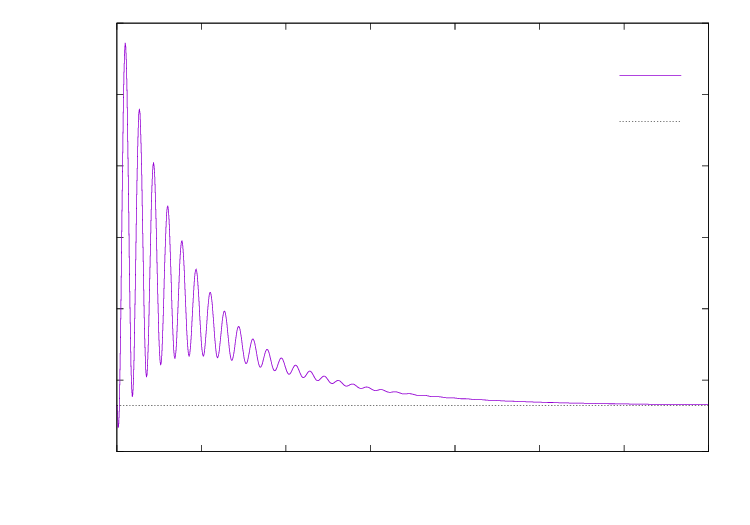}}  \\
       a) & b) 
   \end{tabular}
    \end{center}
    \caption{Stable (3+1)-splitting  solution \eqref{sol.de.sitter}. This orbit is a white point from Figure \ref{fig1}a). In Panel a) it is shown the 3 expanding dimensions together with the stable solution $H=0.5201$ in dotted. Panel b) shows the shrinking dimension with the stable $h=-0.2407$ in dotted. Once $\alpha=0.017$ and $\gamma=-16$ are chosen from Figure \ref{fig5}a), while $H=0.5201$ and $h=-0.2407$ from Figure\ref{fig5}b), $\beta=48.0061$ and  $\Lambda=-0.505859$ are uniquely specified from \eqref{sol.de.sitter}\label{fig2}}
    \end{figure}
     \begin{figure}[htpb]
 \begin{center}
\begin{tabular}{c}
 \resizebox{\imsize}{!}{\input{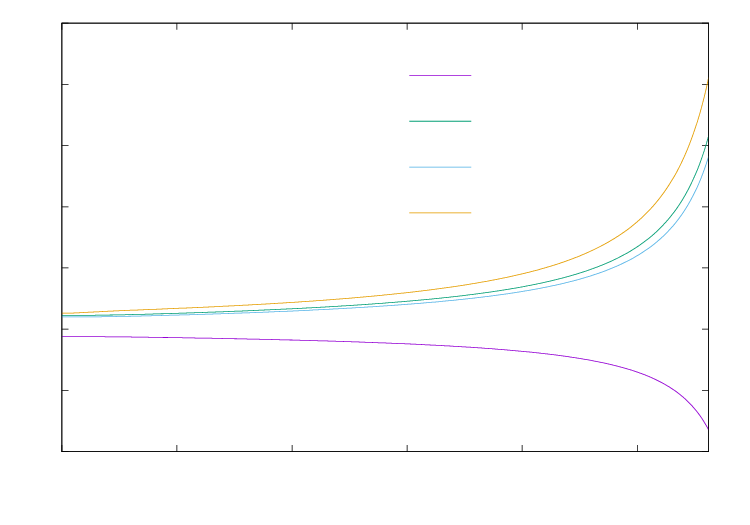}}  
   \end{tabular}
    \end{center}
    \caption{A typical singular orbit. Curvature scalars increase and derivatives of Hubble parameters also increase, which characterizes a curvature singularity. Now this orbit is a black point from Figure \ref{fig1}.\label{fig3}}
    \end{figure}
    \begin{figure}[htpb]
 \begin{center}
\begin{tabular}{c c}
 \resizebox{\halfsize}{!}{\input{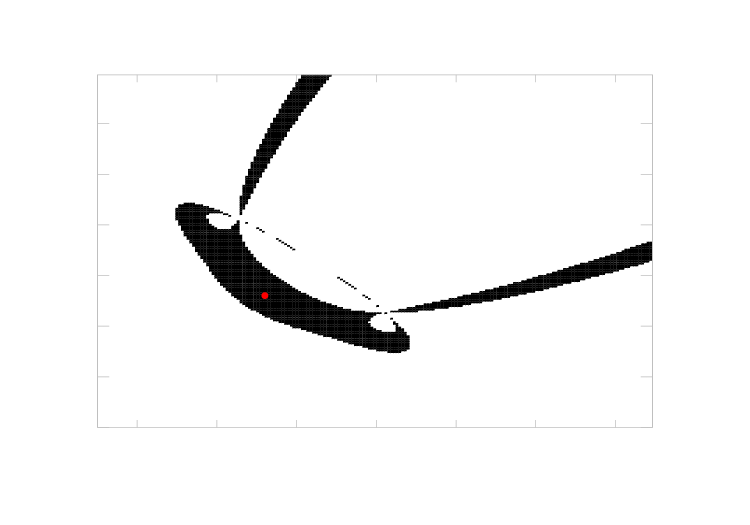}}   & 
 \resizebox{\halfsize}{!}{\input{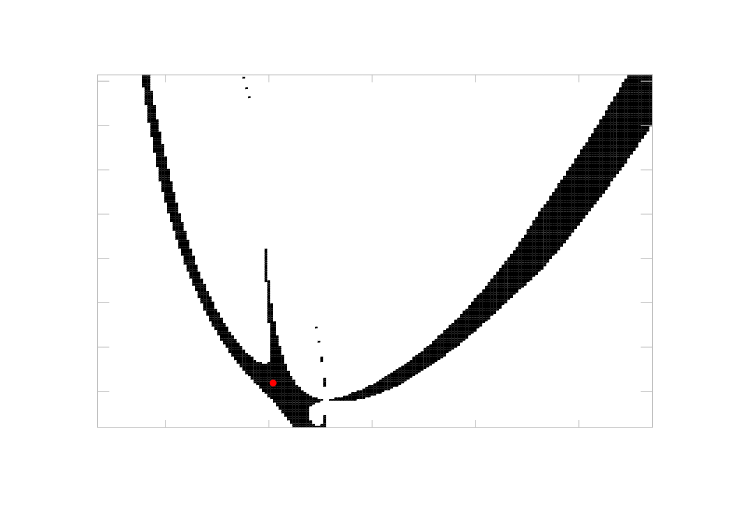}}  \\
       a) & b) 
   \end{tabular}
    \end{center}
    \caption{Basin plot in initial condition space for Hubble parameters. In both plots the coupling constants are fixed to $\alpha=0.017$, $\beta=48.0061$, $\gamma=-16$, $\Lambda=-0.505859$ and each black point corresponds to a ``good'' initial condition, which asymptotes the constant 3+1 splitting solution denoted as the red point, while each white point goes to the singularity. Panel a)The plane $H_1 \times H_2$. Panel b) the plane $H_1\times h_1$. From Panel b) it is possible to see that it is not necessary to have initially one contracting dimension and 3 expanding. Although of rather small measure region, initially sufficient anisotropy can be  enough to reach the (3+1) splitting solution.   \label{fig1}}
    \end{figure}
    For both plots in Figure \ref{fig1} all black points converge to the (3+1) splitting orbit shown in red. This confirms stability of the solution \eqref{sol.de.sitter} for specific value of coupling constants. The (3+1) splitting orbit follows by choosing $\alpha=0.017$ and $\gamma=-16$ from Figure \ref{fig5}a) while $h=-0.2407$ and $H=0.5201$ are chosen from Figure \ref{fig5}b), and then through \eqref{sol.de.sitter}, $\beta=48.0061$  $\Lambda=-0.505859$ are uniquely specified. Figure \ref{fig1}b) shows that although of rather small measure, to obtain the (3+1) splitting it is not necessary that initially one of the dimension is contracting, though  a sufficient anisotropy is required. 
\section{Conclusions}

In the present paper we have considered a possibility for initially fully anisotropic 
4-dim Universe to converge dynamically to a product of isotropic 3-dim Universe and
a 1-dim world in a general quadratic gravity. This scenario could represent a dynamical compactification process needed to "create" our 3-dim world in the framework of multidimensional cosmology.

Such solutions can be found analytically, and the question to solve is their stability. The number
of degrees of freedom in the problem under investigation is rather large, this prevents us from 
finding close formulae for corresponding eigenvalues. We have been forced to use a combination
of analytical approach with numerics.

As a result, we can claim that a zone of coupling constants leading to stable compactification solution exists though being rather restricted. Numerical simulations of the 4-dim Universe evolution show that the dynamics converges to the desired solution in initial conditions belong
to corresponding basin of attractor which in our case represents a narrow (but with a positive
measure) zone in the initial conditions space.

Our results show that a realization  of dynamical compactification scenario in the general 
quadratic gravity is more difficult than in the Gauss-Bonnet cosmology where positivity of 
the volume expansion rate is almost enough for compactification solution to be stable.
On the other hand, it is not impossible, and no severe fine-tuning is needed for such a
scenario to be realized in 4-dim case.

Further study is needed to understand if this scenario is still possible in the general
quadratic gravity for bigger number of spatial dimensions (we remind a reader that stability
results for splitting solutions in the Gauss-Bonnet cosmology do not depend on the number of dimensions).

\subsection*{Acknowledgements}
AT is grateful to Universidad de Brasilia (Brazil) where part of this work have been done, for hospitality. The study was conducted under the state assignment of Lomonosov Moscow State University.

\onecolumn
\section*{Appendix}

\begin{align*}
& \frac{E_{0}}{\sqrt{-g}}=-\frac{\Lambda}{8\pi G}-\frac{H_{3}h_{1}}{8\pi G}-\frac{H_{2}h_{1}}{8\pi G}-\frac{H_{1}h_{1}}{8\pi G}-\frac{H_{2}H_{3}}{8\pi G}-\frac{H_{1}H_{3}}{8\pi G}-\frac{H_{1}H_{2}}{8\pi G}+\beta\left(-4h_{1}\ddot{h}_{1}\right. \nonumber\\
 & -2H_{3}\ddot{h}_{1}-2H_{2}\ddot{h}_{1}+2\dot{H}_{2}\dot{H}_{3}+2\dot{H}_{1}\dot{H}_{3}+2\dot{H}_{2}\dot{h}_{1}+2\dot{H}_{1}\dot{h}_{1}-2H_{2}\ddot{H}_{3}-2H_{1}\ddot{H}_{3}-2H_{1}\ddot{h}_{1}-2h_{1}\ddot{H}_{3} \nonumber\\
 & -2h_{1}\ddot{H}_{2}-2H_{3}\ddot{H}_{2}-4H_{2}\ddot{H}_{2}-2H_{1}\ddot{H}_{2}-2h_{1}\ddot{H}_{1}-2H_{3}\ddot{H}_{1}-2H_{2}\ddot{H}_{1}-4H_{1}\ddot{H}_{1}+2\dot{h}_{1}^{2}+2H_{3}\dot{h}_{1} \nonumber\\
 & -4h_{1}^{2}\dot{h}_{1}-8H_{3}h_{1}\dot{h}_{1}-8H_{2}h_{1}\dot{h}_{1}-8H_{1}h_{1}\dot{h}_{1}-4H_{2}H_{3}\dot{h}_{1}-4H_{1}H_{3}\dot{h}_{1}-4H_{1}H_{2}\dot{h}_{1}+2\dot{H}_{3}^{2}-4H_{3}\ddot{H}_{3} \nonumber\\
 & -8H_{3}h_{1}\dot{H}_{3}-4H_{2}h_{1}\dot{H}_{3}-4H_{1}h_{1}\dot{H}_{3}-4H_{3}^{2}\dot{H}_{3}-8H_{2}H_{3}\dot{H}_{3}-8H_{1}H_{3}\dot{H}_{3}-4H_{1}H_{2}\dot{H}_{3}+2\dot{H}_{2}^{2} \nonumber\\
 & +2\dot{H}_{1}\dot{H}_{2}-4H_{3}h_{1}\dot{H}_{2}-8H_{2}h_{1}\dot{H}_{2}-4H_{1}h_{1}\dot{H}_{2}-8H_{2}H_{3}\dot{H}_{2}-4H_{1}H_{3}\dot{H}_{2}-4H_{2}^{2}\dot{H}_{2}-8H_{1}H_{2}\dot{H}_{2} \nonumber\\
 & -8H_{1}H_{3}\dot{H}_{1}-8H_{1}H_{2}\dot{H}_{1}-4H_{1}^{2}\dot{H}_{1}+4H_{1}^{2}h_{1}^{2}-2H_{2}H_{3}^{2}h_{1}+4H_{2}^{2}H_{3}^{2}-2H_{1}H_{2}H_{3}^{2}-2H_{1}H_{3}^{2}h_{1} \nonumber\\
 & +4H_{1}^{2}H_{3}^{2}-2H_{2}^{3}H_{3}-2H_{1}H_{2}^{2}H_{3}-2H_{1}^{2}H_{2}H_{3}-2H_{1}^{3}H_{3}+2H_{2}^{4}-2H_{1}H_{2}^{3}+4H_{1}^{2}H_{2}^{2}-2H_{1}^{3}H_{2}+2H_{1}^{4} \nonumber\\
 & -2H_{2}^{2}H_{3}h_{1}-2H_{1}^{2}H_{3}h_{1}-2H_{2}^{3}h_{1}-2H_{1}H_{2}^{2}h_{1}-2H_{1}^{2}H_{2}h_{1}-2H_{1}^{3}h_{1}+2H_{3}^{4}-2H_{2}H_{3}^{3}-2H_{1}H_{3}^{3} \nonumber\\
 & -2H_{3}^{3}h_{1}+2h_{1}^{4}-2H_{3}h_{1}^{3}-2H_{2}h_{1}^{3}-2H_{1}h_{1}^{3}+4H_{3}^{2}h_{1}^{2}-2H_{2}H_{3}h_{1}^{2}-2H_{1}H_{3}h_{1}^{2}+4H_{2}^{2}h_{1}^{2}-2H_{1}H_{2}h_{1}^{2} \nonumber\\
 & \left.+2\dot{H}_{1}^{2}-4H_{3}h_{1}\dot{H}_{1}-4H_{2}h_{1}\dot{H}_{1}-8H_{1}h_{1}\dot{H}_{1}-4H_{2}H_{3}\dot{H}_{1}\right)+\alpha\left(12H_{1}^{2}H_{3}^{2}-8H_{2}^{3}H_{3}-8H_{1}^{3}H_{3}\right. \nonumber\\
 & -8H_{2}h_{1}\dot{h}_{1}-8H_{2}H_{3}\dot{H}_{2}-8H_{2}^{2}\dot{H}_{2}-8H_{1}h_{1}^{3}+12H_{3}^{2}h_{1}^{2}+12H_{2}^{2}h_{1}^{2}+4H_{2}^{4}-8H_{1}H_{2}^{3}+12H_{1}^{2}H_{2}^{2} \nonumber\\
 & +12H_{1}^{2}h_{1}^{2}-8H_{3}^{3}h_{1}-8H_{2}^{3}h_{1}-8H_{1}^{3}h_{1}+4H_{3}^{4}-8H_{2}H_{3}^{3}-8H_{1}H_{3}^{3}+12H_{2}^{2}H_{3}^{2}-8H_{1}^{3}H_{2}+4H_{1}^{4} \nonumber\\
 & -8H_{1}H_{2}\dot{H}_{2}+4\dot{H}_{1}^{2}-8H_{1}h_{1}\dot{H}_{1}-8H_{1}H_{3}\dot{H}_{1}-8H_{1}H_{2}\dot{H}_{1}-8H_{1}^{2}\dot{H}_{1}+4h_{1}^{4}-8H_{3}h_{1}^{3}-8H_{2}h_{1}^{3} \nonumber\\
 & -8H_{3}h_{1}\dot{h}_{1}-8H_{1}h_{1}\dot{h}_{1}+4\dot{H}_{3}^{2}-8H_{3}h_{1}\dot{H}_{3}-8H_{3}^{2}\dot{H}_{3}-8H_{2}H_{3}\dot{H}_{3}-8H_{1}H_{3}\dot{H}_{3}+4\dot{H}_{2}^{2}-8H_{2}h_{1}\dot{H}_{2} \nonumber\\
 & \left.-8h_{1}\ddot{h}_{1}-8H_{3}\ddot{H}_{3}-8H_{2}\ddot{H}_{2}-8H_{1}\ddot{H}_{1}+4\dot{h}_{1}^{2}-8h_{1}^{2}\dot{h}_{1}\right)+\gamma\left(-8H_{1}H_{2}H_{3}^{2}+4H_{1}^{2}H_{3}^{2}-8H_{1}H_{2}^{2}H_{3}\right. \nonumber\\
 & -8H_{3}^{2}\dot{H}_{3}-24H_{2}H_{3}\dot{H}_{2}-16H_{1}H_{3}\dot{H}_{2}-24H_{1}H_{3}\dot{H}_{1}-24H_{1}H_{2}\dot{H}_{1}+4H_{1}^{2}H_{2}^{2}-8H_{1}^{2}H_{2}H_{3}+4H_{2}^{4} \nonumber\\
 & +4H_{1}^{4}-8H_{1}H_{3}^{2}h_{1}-8H_{2}^{2}H_{3}h_{1}-24H_{1}H_{2}H_{3}h_{1}-8H_{1}^{2}H_{3}h_{1}-8H_{1}H_{2}^{2}h_{1}-8H_{1}^{2}H_{2}h_{1}+4H_{3}^{4}+4H_{2}^{2}H_{3}^{2} \nonumber\\
 & -8H_{1}^{2}\dot{H}_{1}+4h_{1}^{4}+4H_{3}^{2}h_{1}^{2}-8H_{2}H_{3}h_{1}^{2}-8H_{1}H_{3}h_{1}^{2}+4H_{2}^{2}h_{1}^{2}-8H_{1}H_{2}h_{1}^{2}+4H_{1}^{2}h_{1}^{2}-8H_{2}H_{3}^{2}h_{1} \nonumber\\
 & +8\dot{H}_{2}\dot{H}_{3}-8H_{2}^{2}\dot{H}_{2}-24H_{1}H_{2}\dot{H}_{2}+4\dot{H}_{1}^{2}-16H_{3}h_{1}\dot{H}_{1}-16H_{2}h_{1}\dot{H}_{1}-24H_{1}h_{1}\dot{H}_{1}-16H_{2}H_{3}\dot{H}_{1} \nonumber\\
 & -24H_{1}H_{3}\dot{H}_{3}-16H_{1}H_{2}\dot{H}_{3}+4\dot{H}_{2}^{2}+8\dot{H}_{1}\dot{H}_{2}-16H_{3}h_{1}\dot{H}_{2}-24H_{2}h_{1}\dot{H}_{2}-16H_{1}h_{1}\dot{H}_{2} \nonumber\\
 & -8H_{3}\ddot{h}_{1}-8H_{2}\ddot{h}_{1}-24H_{2}H_{3}h_{1}-8H_{1}\ddot{h}_{1}-8h_{1}\ddot{H}_{3}-16H_{2}H_{3}\dot{h}_{1}-8H_{2}\ddot{H}_{1}-8H_{1}\ddot{H}_{1} \nonumber\\
 & -16H_{1}H_{3}\dot{h}_{1}-16H_{1}H_{2}\dot{h}_{1}+4\dot{H}_{3}^{2}+8\dot{H}_{1}\dot{H}_{3}-24H_{3}h_{1}\dot{H}_{3}-16H_{2}h_{1}\dot{H}_{3}-16H_{1}h_{1}\dot{H}_{3} \nonumber\\
 & +4\dot{h}_{1}^{2}+8\dot{H}_{3}\dot{h}_{1}+8\dot{H}_{2}\dot{h}_{1}+8\dot{H}_{1}\dot{h}_{1}-8h_{1}^{2}\dot{h}_{1}-24H_{3}h_{1}\dot{h}_{1}-24H_{2}h_{1}\dot{h}_{1}-24H_{1}h_{1}\dot{h}_{1} \nonumber\\
 & -8H_{3}\ddot{H}_{3}-8H_{2}\ddot{H}_{3}-8H_{1}\ddot{H}_{3}-8h_{1}\ddot{H}_{2}-8H_{3}\ddot{H}_{2}-8H_{2}\ddot{H}_{2}-8H_{1}\ddot{H}_{2}-8h_{1}\ddot{H}_{1}-8H_{3}\ddot{H}_{1} \nonumber\\
 & \left.-8h_{1}\ddot{h}_{1}\right)\nonumber\\
 \end{align*}
 \begin{align*}
   & \frac{E_{1}}{\sqrt{-g}}=(8\gamma+4\beta+8\alpha)\dddot{H}_{1}+(8\gamma+2\beta)\dddot{H}_{2}+(8\gamma+2\beta)\dddot{H}_{3}+(8\gamma+2\beta)\dddot{h}_{1}+\frac{\Lambda}{8\pi G}+\frac{\dot{h}_{1}}{8\pi G}\\
   &+\frac{\dot{H}_{3}}{8\pi G}+\frac{\dot{H}_{2}}{8\pi G}+\frac{h_{1}^{2}}{8\pi G}+\frac{H_{3}h_{1}}{8\pi G}+\frac{H_{2}h_{1}}{8\pi G}+\frac{H_{3}^{2}}{8\pi G}+\frac{H_{2}H_{3}}{8\pi G}+\frac{H_{2}^{2}}{8\pi G}+24\gamma h_{1}\ddot{h}_{1}+6\beta h_{1}\ddot{h}_{1}\\
   &+16H_{2}\gamma\ddot{h}_{1}+8H_{1}\gamma\ddot{h}_{1}+4\beta H_{3}\ddot{h}_{1}+4H_{2}\beta\ddot{h}_{1}+2H_{1}\beta\ddot{h}_{1}+16H_{3}\gamma\ddot{h}_{1}+16\gamma h_{1}\ddot{H}_{3}+4\beta h_{1}\ddot{H}_{3}\\
   &+24H_{3}\gamma\ddot{H}_{3}+16H_{2}\gamma\ddot{H}_{3}+8H_{1}\gamma\ddot{H}_{3}+6\beta H_{3}\ddot{H}_{3}+16H_{3}\gamma\ddot{H}_{2}+24H_{2}\gamma\ddot{H}_{2}+4\beta h_{1}\ddot{H}_{2}\\
   &+4H_{2}\beta\ddot{H}_{3}+2H_{1}\beta\ddot{H}_{3}+16\gamma h_{1}\ddot{H}_{2}+2H_{1}\beta\ddot{H}_{2}+16\gamma h_{1}\ddot{H}_{1}+8\beta h_{1}\ddot{H}_{1}+16\alpha h_{1}\ddot{H}_{1}\\
   &+8H_{1}\gamma\ddot{H}_{2}+4\beta H_{3}\ddot{H}_{2}+6H_{2}\beta\ddot{H}_{2}+16H_{2}\gamma \ddot{H}_{1} +16H_{1}\gamma\ddot{H}_{1}+8\beta H_{3}\ddot{H}_{1}+16\alpha H_{3}\ddot{H}_{1}+4\alpha\dot{h}_{1}^{2}\\
   &+16H_{3}\gamma\ddot{H}_{1}+8H_{1}\beta\ddot{H}_{1}+16\alpha H_{2}\ddot{H}_{1}+16H_{1}\alpha\ddot{H}_{1}+20\gamma\dot{h}_{1}^{2}+6\beta\dot{h}_{1}^{2}+24\gamma\dot{H}_{3}\dot{h}_{1}+24\gamma h_{1}^{2}\dot{h}_{1}\\
   &+8H_{1}\beta\ddot{H}_{1}+16\alpha H_{2} \ddot{H}_{1}+16H_{1}\alpha\ddot{H}_{1}+20\gamma\dot{h}_{1}^{2}+6\beta\dot{h}_{1}^{2}+4\alpha\dot{h}_{1}^{2}+24\gamma\dot{H}_{3}\dot{h}_{1}+8H_{2}\beta\ddot{H}_{1}\\
&+6\beta\dot{H}_{3}\dot{h}_{1}+24\gamma\dot{H}_{2}\dot{h}_{1}+6\beta\dot{H}_{2}\dot{h}_{1}+16\gamma\dot{H}_{1}\dot{h}_{1} +6\beta\dot{H}_{1}\dot{h}_{1}+8\alpha\dot{H}_{1}\dot{h}_{1}+8\beta H_{3}h_{1}\dot{h}_{1}\\&+8H_{2}\beta h_{1}\dot{h}_{1}-2H_{1}\beta h_{1}\dot{h}_{1}-16H_{1}\alpha h_{1}\dot{h}_{1}+16H_{3}^{2}\gamma\dot{h}_{1}+24H_{2}H_{3}\gamma\dot{h}_{1}+8H_{1}H_{3}\gamma\dot{h}_{1}\\&+16H_{2}^{2}\gamma\dot{h}_{1}+8H_{1}H_{2}\gamma\dot{h}_{1}+4\beta H_{3}^{2}\dot{h}_{1}+4H_{2}\beta H_{3}\dot{h}_{1}+2H_1 \beta H_{3}\dot{h}_{1}+4H_{2}^{2}\beta\dot{h}_{1}+8H_{1}\gamma h_{1}\dot{h}_{1}\\&+2H_{1}H_{2}\beta\dot{h}_{1}+2H_{1}^{2}\beta\dot{h}_{1}+8H_{1}^{2}\alpha\dot{h}_{1}+20\gamma \dot{H}_{3}^{2}+6\beta\dot{H}_{3}^{2}+4\alpha\dot{H}_{3}^{2}+24\gamma\dot{H}_{2}\dot{H}_{3}+8\alpha h_{1}^{2}\dot{h}_{1}\\&+6\beta\dot{H}_{2}\dot{H}_{3}+16\gamma\dot{H}_{1}\dot{H}_{3}+6\beta\dot{H}_{1}\dot{H}_{3}+8\alpha\dot{H}_{1}\dot{H}_{3}+16\gamma h_{1}^{2}\dot{H}_{3}+4\beta h_{1} ^{2}\dot{H}_{3}+32H_{2}\gamma h_{1}\dot{h}_{1}\\&+32H_{3}\gamma h_{1}\dot{H}_{3}+24H_{2}\gamma h_{1}\dot{H}_{3}+8H_{1}\gamma h_{1}\dot{H}_{3}+8\beta H_{3}h_{1}\dot{H}_{3}+4H_{2}\beta h_{1}\dot{H}_{3}+32H_{3}\gamma h_{1}\dot{h}_{1}\\&+2H_{1}\beta h_{1}\dot{H}_{3}+24 H_{3} ^{2}\gamma\dot{H}_{3}+32H_{2}H_{3}\gamma\dot{H}_{3}+8H_{1}H_{3}\gamma \dot{H}_{3}+16H_{2}^{2}\gamma\dot{H}_{3}+8\beta h_{1}^{2}\dot{h}_{1}+4\alpha\dot{H}_{2}^{2}\\&+8H_{1}H_{2}\gamma\dot{H}_{3}+8\beta H_{3}^{2}\dot{H}_{3}+8\alpha H_{3}^{2}\dot{H}_{3}+8H_{2}\beta H_{3} \dot{H}_{3}-2H_{1}\beta H_{3}\dot{H}_{3}-16H_{1}\alpha H_{3}\dot{H}_{3}+6\beta\dot{H}_{2}^{2}\\&+4H_{2}^{2}\beta\dot{H}_{3}+2H_{1}H_{2}\beta\dot{H}_{3}+2H_{1}^{2}\beta\dot{H}_{3}+8H_{1}^{2}\alpha\dot{H}_{3}+20\gamma\dot{H}_{2}^{2}+4\beta h_{1} ^{2}\dot{H}_{2}+24H_{3}\gamma h_{1}\dot{H}_{2}\\&+32H_{2}\gamma h_{1}\dot{H}_{2}+8H_{1}\gamma h_{1}\dot{H}_{2}+4\beta H_{3}h_{1}\dot{H}_{2}+8H_{2}\beta h_{1}\dot{H}_{2}+16\gamma\dot{H}_{1}\dot{H}_{2}+6\beta\dot{H}_{1}\dot{H}_{2}\\&+16H_{3} ^{2}\gamma\dot{H}_{2}+32H_{2}H_{3}\gamma\dot{H}_{2}+8H_{1}H_{3}\gamma \dot{H}_{2}+24H_{2}^{2}\gamma\dot{H}_{2}+8H_{1}H_{2}\gamma\dot{H}_{2}+4\beta H_{3}^{2}\dot{H}_{2}\\&+8H_{2}\beta H_{3}\dot{H}_{2}+2H_{1}\beta H_{3}\dot{H}_{2}+8H_{2}^{2}\beta\dot{H}_{2}-2H_{1}H_{2}\beta \dot{H}_{2} +2H_{1}^{2}\beta\dot{H}_{2}+8\alpha H_{2}^{2}\dot{H}_{2}+16\gamma h_{1}^{2}\dot{H}_{2}\\&-16H_{1}\alpha H_{2}\dot{H}_{2}+8H_{1}^{2}\alpha\dot{H}_{2}+12\gamma\dot{H}_{1}^{2}+6\beta \dot{H}_{1}^{2}+12\alpha\dot{H}_{1}^{2}+8\gamma h_{1}^{2}\dot{H}_{1}+2\beta h_{1}^{2}\dot{H}_{1}+8\alpha\dot{H}_{1}\dot{H}_{2}\\&+16H_{3}\gamma h_{1}\dot{H}_{1}+16H_{2}\gamma h_{1} \dot{H}_{1}+8H_{1}\gamma h_{1}\dot{H}_{1}+8\beta H_{3}h_{1}\dot{H}_{1}+16\alpha H_{3}h_{1}\dot{H}_{1}+2H_{1}\beta h_{1}\dot{H}_{2}\\&+8H_{2}\beta h_{1}\dot{H}_{1}+10H_1 \beta h_{1}\dot{H}_{1}+16\alpha H_{2}h_{1}\dot{H}_{1}+32H_{1}\alpha h_{1}\dot{H}_{1}+8H_{3}^{2}\gamma\dot{H}_{1}-8H_{1}^{2}\gamma\dot{H}_{1}\\&+8H_{2}\beta H_{3}\dot{H}_{1}+10H_{1}\beta H_{3}\dot{H}_{1}+16\alpha H_{2}H_{3}\dot{H}_{1}+32H_{1}\alpha H_{3}\dot{H}_{1}+2H_{2} ^{2}\beta\dot{H}_{1}-4H_{1}^{2}\beta \dot{H}_{1}\\&+10H_{1}H_{2}\beta\dot{H}_{1}+32H_{1}\alpha H_{2}\dot{H}_{1}-8H_{1}^{2}\alpha\dot{H}_{1}+4\gamma h_{1}^{4}+2\beta h_{1}^{4}+4\alpha h_{1}^{4}+8H_{3}\gamma h_{1}^{3}+8H_{2}\gamma h_{1}^{3}\\&+2\beta H_{3}h_{1}^{3}+2H_{2}\beta h_{1}^{3}-2H_{1}\beta h_{1}^{3}-8H_{1}\alpha h_{1}^{3}+12H_{3}^{2}\gamma h_{1}^{2}+16H_{2}H_{3}\gamma h_{1}^{2}+12H_{2}^{2}\gamma h_{1}^{2}+8H_{2}^{2}\gamma\dot{H}_{1}\\&-4H_{1}^{2}\gamma h_{1}^{2}+4\beta H_{3}^{2}h_{1} ^{2}+4\alpha H_{3}^{2}h_{1}^{2}+2H_{2}\beta H_{3}h_{1}^{2}-2H_1 \beta H_{3}h_{1}^{2}-8H_{1}\alpha H_{3}h_{1}^{2}+4H_{2}^{2}\beta h_{1}^{2}\\&-2H_{1}H_{2}\beta h_{1}^{2}+4\alpha H_{2}^{2}h_{1}^{2}-8H_{1}\alpha H_{2}h_{1}^{2}+4H_{1}^{2}\alpha h_{1}^{2}+8H_{3}^{3}\gamma h_{1}+16H_{2}H_{3}^{2}\gamma h_{1}+16H_{2}^{2}H_{3}\gamma h_{1}\\&-8H_{1}^{2}H_{3}\gamma h_{1}+8H_{2}^{3}\gamma h_{1}-8H_{1}^{2}H_{2}\gamma h_{1}-8H_{1}^{3}\gamma h_{1} +2\beta H_{3}^{3}h_{1}+2H_{2}\beta H_{3}^{2}h_{1}-2H_{1}\beta H_{3}^{2}h_{1}\\&-8H_{1}\alpha H_{3}^{2}h_{1}+2H_{2}^{2}\beta H_{3}h_{1}+2H_{1}^{2}\beta H_{3}h_{1}+16H_{1}^{2}\alpha H_{3}h_{1}+2H_{2}^{3}\beta h_{1}-2H_{1}H_{2}^{2}\beta h_{1}+2\beta H_{3}^{2}\dot{H}_{1}\\&+2H_{1}^{2}H_{2}\beta h_{1}-2H_{1}^{3}\beta h_{1}-8H_1 \alpha H_{2}^{2}h_{1}+16H_{1}^{2}\alpha H_{2}h_{1}+4H_{3} ^{4}\gamma+8H_{2}H_{3}^{3}\gamma+12H_{2}^{2}H_{3}^{2}\gamma-4H_{1}^{4}\gamma\\&-4H_{1}^{2}H_{3}^{2}\gamma+8H_{2}^{3}H_{3}\gamma-8H_{1}^{2}H_{2}H_{3} \gamma-8H_{1}^{3}H_{3}\gamma+4H_{2}^{4}\gamma-4H_{1}^{2}H_{2} ^{2}\gamma-8H_{1}^{3}H_{2}\gamma+8H_{1}H_{2}\gamma\dot{H}_{1}\\&+2\beta H_{3} ^{4}+4\alpha H_{3}^{4}+2H_{2}\beta H_{3}^{3}-2H_{1}\beta H_{3} ^{3}-8H_{1}\alpha H_{3}^{3}+4H_{2}^{2}\beta H_{3}^{2}-2H_{1}H_{2}\beta H_{3}^{2}+4\alpha H_{2}^{2}H_{3}^{2}\\&-8H_{1}\alpha H_{2}H_{3}^{2}+4H_{1}^{2}\alpha H_{3}^{2}+2H_{2}^{3}\beta H_{3}-2H_{1}H_{2}^{2}\beta H_{3}+2H_{1}^{2}H_{2}\beta H_{3}-2H_{1}^{3}\beta H_{3}-8H_{1}\alpha H_{2}^{2}H_{3}\\&+16H_{1}^{2}\alpha H_{2}H_{3}+2H_{2}^{4}\beta-2H_{1}H_{2}^{3}\beta-2H_{1}^{3}H_{2}\beta-2H_{1}^{4}\beta+4\alpha H_{2}^{4}-8H_{1}\alpha H_{2}^{3}+4H_{1}^{2}\alpha H_{2}^{2}-4H_{1}^{4}\alpha\\&+16H_{2}H_{3}\gamma\dot{H}_{1}+8H_{1}H_{3}\gamma\dot{H}_{1}
\end{align*}
\begin{align*}
   & \frac{E_{2}}{\sqrt{-g}}=(8\gamma+2\beta)\dddot{H}_{1}+(8\gamma+4\beta+8\alpha)\dddot{H}_{2}+(8\gamma+2\beta)\dddot{H}_{3}+(8\gamma+2\beta)\dddot{h}_{1}+\frac{\Lambda}{8\pi G}+\frac{\dot{h}_{1}}{8\pi G}\\&+\frac{\dot{H}_{3}}{8\pi G}+\frac{\dot{H}_{1}}{8\pi G}+\frac{h_{1}^{2}}{8\pi G}+\frac{H_{3}h_{1}}{8\pi G}+\frac{H_{1}h_{1}}{8\pi G}+\frac{H_{3}^{2}}{8\pi G}+\frac{H_{1}H_{3}}{8\pi G}+\frac{H_{1}^{2}}{8\pi G}+24\gamma h_{1}\ddot{h}_{1}+6\beta h_{1}\ddot{h}_{1}\\&+16H_{3}\gamma\ddot{h}_{1}+8H_{2}\gamma\ddot{h}_{1}+16H_{1}\gamma\ddot{h}_{1}+4\beta H_{3}\ddot{h}_{1}+2H_{2}\beta\ddot{h}_{1}+4H_{1}\beta\ddot{h}_{1}+16\gamma h_{1}\ddot{H}_3+4\beta h_{1}\ddot{H}_{3}\\&+24H_{3}\gamma\ddot{H}_{3}+8H_{2}\gamma\ddot{H}_{3}+16H_{1}\gamma\ddot{H}_{3}+6\beta H_{3}\ddot{H}_{3}+2H_{2}\beta \ddot{H}_{3}+4H_{1}\beta\ddot{H}_{3}+16\gamma h_{1}\ddot{H}_{2}\\&+8\beta h_{1}\ddot{H}_{2}+16\alpha h_{1}\ddot{H}_{2}+16H_{3}\gamma\ddot{H}_{2}+16H_{2}\gamma\ddot{H}_{2}+16H_{1}\gamma\ddot{H}_{2}+8\beta H_{3}\ddot{H}_{2}+16\alpha H_{3}\ddot{H}_{2}\\&+8H_{2}\beta\ddot{H}_{2}+8H_{1}\beta \ddot{H}_2+16\alpha H_{2}\ddot{H}_{2}+16H_{1}\alpha\ddot{H}_{2}+16\gamma h_{1}\ddot{H}_{1}+4\beta h_{1}\ddot{H}_{1}+16H_{3}\gamma\ddot{H}_{1}+8H_{2}\gamma\ddot{H}_{1}\\&+24H_{1}\gamma\ddot{H}_{1}+4\beta H_{3}\ddot{H}_{1}+2H_{2}\beta \ddot{H}_{1}+6H_{1}\beta\ddot{H}_{1}+20\gamma\dot{h}_{1}^{2}+6\beta\dot{h}_{1}^{2}+4\alpha\dot{h}_{1}^{2}+24\gamma\dot{H}_{3}\dot{h}_{1}+6\beta\dot{H}_{3}\dot{h}_{1}\\&+16\gamma\dot{H}_{2}\dot{h}_{1}+6\beta\dot{H}_{2}\dot{h}_{1}+8\alpha\dot{H}_{2}\dot{h}_1+24\gamma\dot{H}_{1}\dot{h}_{1}+6\beta\dot{H}_{1}\dot{h}_{1}+24\gamma h_{1}^{2}\dot{h}_{1}+8\beta h_{1}^{2}\dot{h}_{1}+8\alpha h_{1}^{2}\dot{h}_{1}\\&+32H_{3}\gamma h_{1}\dot{h}_{1}+8H_{2}\gamma h_{1}\dot{h}_{1}+32H_{1}\gamma h_{1}\dot{h}_{1}+8\beta H_{3}h_{1}\dot{h}_{1}-2H_{2}\beta h_{1}\dot{h}_{1}+8H_{1}\beta h_{1}\dot{h}_{1}-16\alpha H_{2}h_{1}\dot{h}_{1}\\&+16H_{3}^{2}\gamma\dot{h}_{1}+8H_{2}H_{3}\gamma\dot{h}_{1}+24H_{1}H_{3}\gamma\dot{h}_{1}+8H_{1}H_{2}\gamma\dot{h}_{1}+16H_{1}^{2}\gamma\dot{h}_{1}+4\beta H_{3}^{2}\dot{h}_{1}+2H_{2}\beta H_{3}\dot{h}_{1}\\&+4H_1\beta H_{3}\dot{h}_{1}+2H_{2}^{2}\beta\dot{h}_{1}+2H_{1}H_{2}\beta\dot{h}_{1}+4H_{1}^{2}\beta\dot{h}_{1}+8\alpha H_{2}^{2}\dot{h}_{1}+20\gamma \dot{H}_{3}^{2}+6\beta\dot{H}_{3}^{2}+4\alpha\dot{H}_{3}^{2}\\&+16\gamma\dot{H}_{2}\dot{H}_{3}+6\beta\dot{H}_{2}\dot{H}_{3}+8\alpha\dot{H}_{2}\dot{H}_{3}+24\gamma\dot{H}_{1}\dot{H}_{3}+6\beta\dot{H}_{1}\dot{H}_{3}+16\gamma h_{1}^{2}\dot{H}_{3}+4\beta h_1^{2}\dot{H}_{3}\\&+32H_{3}\gamma h_{1}\dot{H}_{3}+8H_{2}\gamma h_{1}\dot{H}_{3}+24H_{1}\gamma h_{1}\dot{H}_{3}+8\beta H_{3}h_{1}\dot{H}_{3}+2H_{2}\beta h_{1}\dot{H}_{3}+4H_{1}\beta h_{1}\dot{H}_{3}+24H_3^{2}\gamma\dot{H}_{3}\\&+8H_{2}H_{3}\gamma\dot{H}_{3}+32H_{1}H_{3}\gamma \dot{H}_{3}+8H_{1}H_{2}\gamma\dot{H}_{3}+16H_{1}^{2}\gamma\dot{H}_{3}+8\beta H_{3}^{2}\dot{H}_{3}+8\alpha H_{3}^{2}\dot{H}_{3}-2H_{2}\beta H_3\dot{H}_{3}\\&+8H_{1}\beta H_{3}\dot{H}_{3}-16\alpha H_{2}H_{3}\dot{H}_3+2H_{2}^{2}\beta\dot{H}_{3}+2H_{1}H_{2}\beta\dot{H}_{3}+4H_{1}^{2}\beta\dot{H}_{3}+8\alpha H_{2}^{2}\dot{H}_{3}+12\gamma\dot{H}_{2}^{2}+6\beta \dot{H}_{2}^{2}\\&+12\alpha\dot{H}_{2}^{2}+16\gamma\dot{H}_{1}\dot{H}_{2}+6\beta \dot{H}_1\dot{H}_{2}+8\alpha\dot{H}_{1}\dot{H}_{2}+8\gamma h_{1}^{2}\dot{H}_{2}+2\beta h_1^{2}\dot{H}_{2}+16H_{3}\gamma h_{1}\dot{H}_{2}+8H_{2}\gamma h_{1}\dot{H}_{2}\\&+16H_{1}\gamma h_{1}\dot{H}_{2}+8\beta H_{3}h_{1}\dot{H}_{2}+16\alpha H_{3}h_{1}\dot{H}_{2}+10H_{2}\beta h_{1}\dot{H}_{2}+8H_1\beta h_{1}\dot{H}_{2}+32\alpha H_{2}h_{1}\dot{H}_{2}\\&+16H_{1}\alpha h_{1}\dot{H}_{2}+8H_{3}^{2}\gamma\dot{H}_{2}+8H_{2}H_{3}\gamma\dot{H}_{2}+16H_{1}H_{3}\gamma\dot{H}_{2}-8H_{2}^{2}\gamma\dot{H}_{2}+8H_1H_{2}\gamma\dot{H}_{2}+8H_{1}^{2}\gamma\dot{H}_{2}\\&+2\beta H_{3}^{2}\dot{H}_{2}+10H_{2}\beta H_{3}\dot{H}_{2}+8H_{1}\beta H_{3}\dot{H}_{2}+32\alpha H_{2}H_{3}\dot{H}_{2}+16H_{1}\alpha H_{3}\dot{H}_{2}-4H_2^{2}\beta\dot{H}_{2}\\&+10H_{1}H_{2}\beta\dot{H}_{2}+2H_{1}^{2}\beta \dot{H}_{2}-8\alpha H_{2}^{2}\dot{H}_{2}+32H_{1}\alpha H_{2}\dot{H}_{2}+20\gamma\dot{H}_{1}^{2}+6\beta\dot{H}_{1}^{2}+4\alpha\dot{H}_{1}^{2}\\&+16\gamma h_1^{2}\dot{H}_{1}+4\beta h_{1}^{2}\dot{H}_{1}+24H_{3}\gamma h_{1}\dot{H}_{1}+8H_{2}\gamma h_{1}\dot{H}_{1}+32H_{1}\gamma h_{1}\dot{H}_{1}+4\beta H_{3}h_{1}\dot{H}_{1}+2H_{2}\beta h_{1}\dot{H}_{1}\\&+8H_{1}\beta h_{1}\dot{H}_{1}+16H_{3}^{2}\gamma\dot{H}_{1}+8H_{2}H_{3}\gamma \dot{H}_1+32H_{1}H_{3}\gamma\dot{H}_{1}+8H_{1}H_{2}\gamma\dot{H}_{1}+24H_{1}^{2}\gamma\dot{H}_{1}+4\beta H_{3}^{2}\dot{H}_{1}\\&+2H_{2}\beta H_3\dot{H}_{1}+8H_{1}\beta H_{3}\dot{H}_{1}+2H_{2}^{2}\beta\dot{H}_{1}-2H_1H_{2}\beta\dot{H}_{1}+8H_{1}^{2}\beta\dot{H}_{1}+8\alpha H_{2}^{2}\dot{H}_{1}-16H_{1}\alpha H_{2}\dot{H}_{1}\\&+8H_{1}^{2}\alpha\dot{H}_{1}+4\gamma h_{1}^{4}+2\beta h_{1}^{4}+4\alpha h_{1}^{4}+8H_{3}\gamma h_1^{3}+8H_{1}\gamma h_{1}^{3}+2\beta H_{3}h_{1}^{3}-2H_{2}\beta h_{1}^{3}+2H_{1}\beta h_{1}^{3}\\&-8\alpha H_{2}h_{1}^{3}+12H_{3}^{2}\gamma h_{1}^{2}+16H_{1}H_{3}\gamma h_{1}^{2}-4H_{2}^{2}\gamma h_{1}^{2}+12H_{1}^{2}\gamma h_{1}^{2}+4\beta H_{3}^{2}h_1^{2}+4\alpha H_{3}^{2}h_{1}^{2}-2H_{2}\beta H_{3}h_{1}^{2}\\&+2H_1\beta H_{3}h_{1}^{2}-8\alpha H_{2}H_{3}h_{1}^{2}-2H_{1}H_{2}\beta h_{1}^{2}+4H_{1}^{2}\beta h_{1}^{2}+4\alpha H_{2}^{2}h_{1}^{2}-8H_{1}\alpha H_{2}h_{1}^{2}+4H_{1}^{2}\alpha h_{1}^{2}+8H_{3}^{3}\gamma h_{1}\\&+16H_{1}H_{3}^{2}\gamma h_{1}-8H_{2}^{2}H_{3}\gamma h_{1}+16H_{1}^{2}H_{3}\gamma h_{1}-8H_{2}^{3}\gamma h_{1}-8H_{1}H_{2}^{2}\gamma h_{1}+8H_{1}^{3}\gamma h_1+2\beta H_{3}^{3}h_{1}\\&-2H_{2}\beta H_{3}^{2}h_{1}+2H_{1}\beta H_{3}^{2}h_{1}-8\alpha H_{2}H_{3}^{2}h_{1}+2H_{2}^{2}\beta H_{3}h_{1}+2H_{1}^{2}\beta H_{3}h_{1}+16\alpha H_{2}^{2}H_{3}h_{1}-2H_{2}^{3}\beta h_{1}\\&+2H_{1}H_{2}^{2}\beta h_{1}-2H_{1}^{2}H_{2}\beta h_{1}+2H_{1}^{3}\beta h_{1}+16H_1\alpha H_{2}^{2}h_{1}-8H_{1}^{2}\alpha H_{2}h_{1}+4H_{3}^{4}\gamma+8H_{1}H_{3}^{3}\gamma-4H_{2}^{2}H_{3}^{2}\gamma\\&+12H_1^{2}H_{3}^{2}\gamma-8H_{2}^{3}H_{3}\gamma-8H_{1}H_{2}^{2}H_3\gamma+8H_{1}^{3}H_{3}\gamma-4H_{2}^{4}\gamma-8H_{1}H_2^{3}\gamma-4H_{1}^{2}H_{2}^{2}\gamma+4H_{1}^{4}\gamma+2\beta H_3^{4}\\&+4\alpha H_{3}^{4}-2H_{2}\beta H_{3}^{3}+2H_{1}\beta H_3^{3}-8\alpha H_{2}H_{3}^{3}-2H_{1}H_{2}\beta H_{3}^{2}+4H_1^{2}\beta H_{3}^{2}+4\alpha H_{2}^{2}H_{3}^{2}-8H_{1}\alpha H_2H_{3}^{2}\\&+4H_{1}^{2}\alpha H_{3}^{2}-2H_{2}^{3}\beta H_{3}+2H_{1}H_{2}^{2}\beta H_{3}-2H_{1}^{2}H_{2}\beta H_{3}+2H_1^{3}\beta H_{3}+16H_{1}\alpha H_{2}^{2}H_{3}-8H_{1}^{2}\alpha H_{2}H_{3}\\&-2H_{2}^{4}\beta-2H_{1}H_{2}^{3}\beta-2H_1^{3}H_{2}\beta+2H_{1}^{4}\beta-4\alpha H_{2}^{4}+4H_{1}^{2}\alpha H_{2}^{2}-8H_{1}^{3}\alpha H_{2}+4H_{1}^{4}\alpha
\end{align*}
\begin{align*}
&\frac{E_{3}}{\sqrt{-g}}=(8\gamma+2\beta)\dddot{H}_{1}+(8\gamma+2\beta)\dddot{H}_{2}+(8\gamma+4\beta+8\alpha)\dddot{H}_{3}+(8\gamma+2\beta)\dddot{h}_{1}+\frac{\Lambda}{8\pi G}+\frac{\dot{h}_{1}}{8\pi G}\\&+\frac{\dot{H}_{2}}{8\pi G}+\frac{\dot{H}_{1}}{8\pi G}+\frac{h_{1}^{2}}{8\pi G}+\frac{H_{2}h_{1}}{8\pi G}+\frac{H_{1}h_{1}}{8\pi G}+\frac{H_{2}^{2}}{8\pi G}+\frac{H_{1}H_{2}}{8\pi G}+\frac{H_{1}^{2}}{8\pi G}++24\gamma h_{1}\ddot{h}_{1}+6\beta h_{1}\ddot{h}_{1}\\&+8H_{3}\gamma\ddot{h}_{1}+16H_{2}\gamma\ddot{h}_{1}+16H_{1}\gamma\ddot{h}_{1}+2\beta H_{3}\ddot{h}_{1}+4H_{2}\beta\ddot{h}_{1}+4H_{1}\beta\ddot{h}_{1}+16\gamma h_{1}\ddot{H}_3+8\beta h_{1}\ddot{H}_{3}\\&+16\alpha h_{1}\ddot{H}_{3}+16H_{3}\gamma\ddot{H}_{3}+16H_{2}\gamma\ddot{H}_{3}+16H_{1}\gamma\ddot{H}_{3}+8\beta H_{3}\ddot{H}_{3}+16\alpha H_{3}\ddot{H}_{3}+8H_{2}\beta\ddot{H}_{3}+8H_{1}\beta\ddot{H}_{3}\\&+16\alpha H_{2}\ddot{H}_{3}+16H_{1}\alpha \ddot{H}_3+16\gamma h_{1}\ddot{H}_{2}+4\beta h_{1}\ddot{H}_{2}+8H_{3}\gamma\ddot{H}_{2}+24H_{2}\gamma\ddot{H}_{2}+16H_{1}\gamma\ddot{H}_{2}+2\beta  H_3\ddot{H}_{2}\\&+6H_{2}\beta\ddot{H}_{2}+4H_{1}\beta\ddot{H}_{2}+16\gamma h_1\ddot{H}_{1}+4\beta h_{1}\ddot{H}_{1}+8H_{3}\gamma\ddot{H}_{1}+16H_{2}\gamma\ddot{H}_{1}+24H_{1}\gamma\ddot{H}_{1}+2\beta H_{3}\ddot{H}_{1}\\&+4 H_2\beta\ddot{H}_{1}+6H_{1}\beta\ddot{H}_{1}+20\gamma\dot{h}_{1}^{2}+6\beta\dot{h}_{1}^{2}+4\alpha\dot{h}_{1}^{2}+16\gamma\dot{H}_{3}\dot{h}_{1}+6\beta\dot{H}_{3}\dot{h}_{1}+8\alpha\dot{H}_{3}\dot{h}_{1}+24\gamma\dot{H}_{2}\dot{h}_{1}\\&+6\beta \dot{H}_2\dot{h}_{1}+24\gamma\dot{H}_{1}\dot{h}_{1}+6\beta\dot{H}_{1}\dot{h}_{1}+24\gamma h_{1}^{2}\dot{h}_{1}+8\beta h_{1}^{2}\dot{h}_{1}+8\alpha h_{1}^{2}\dot{h}_{1}+8H_{3}\gamma h_{1}\dot{h}_{1}+32H_{2}\gamma h_{1}\dot{h}_{1}\\&+32H_{1}\gamma h_{1}\dot{h}_{1}-2\beta H_{3}h_{1}\dot{h}_{1}-16\alpha H_{3}h_{1}\dot{h}_{1}+8H_{2}\beta h_{1}\dot{h}_{1}+8H_{1}\beta h_{1}\dot{h}_{1}+8H_{2}H_{3}\gamma\dot{h}_{1}+8H_{1}H_{3}\gamma\dot{h}_{1}\\&+16H_{2}^{2}\gamma\dot{h}_{1}+24H_{1}H_{2}\gamma\dot{h}_{1}+16H_{1}^{2}\gamma\dot{h}_{1}+2\beta H_{3}^{2}\dot{h}_{1}+8\alpha H_{3}^{2}\dot{h}_{1}+2H_{2}\beta H_{3}\dot{h}_{1}+2H_{1}\beta H_{3}\dot{h}_{1}+4H_{2}^{2}\beta\dot{h}_{1}\\&+4H_{1}H_{2}\beta\dot{h}_{1}+4H_{1}^{2}\beta \dot{h}_1+12\gamma\dot{H}_{3}^{2}+6\beta\dot{H}_{3}^{2}+12\alpha\dot{H}_{3}^{2}+16\gamma\dot{H}_{2}\dot{H}_{3}+6\beta\dot{H}_{2}\dot{H}_{3}+8\alpha\dot{H}_{2}\dot{H}_{3}+4\alpha h_{1}^{4}\\&+16 \gamma \dot{H}_{1}\dot{H}_{3}+6\beta\dot{H}_{1}\dot{H}_{3}+8\alpha\dot{H}_{1}\dot{H}_{3}+8\gamma h_{1}^{2}\dot{H}_{3}+2\beta h_{1}^{2}\dot{H}_{3}+8H_{3}\gamma h_{1}\dot{H}_{3}+16H_{2}\gamma h_{1}\dot{H}_{3}\\&+16H_{1}\gamma h_{1}\dot{H}_3+10\beta H_{3}h_{1}\dot{H}_{3}+32\alpha H_{3}h_{1}\dot{H}_{3}+8 H_2\beta h_{1}\dot{H}_{3}+8H_{1}\beta h_{1}\dot{H}_{3}+16\alpha  H_2h_{1}\dot{H}_{3}\\&-8H_{3}^{2}\gamma\dot{H}_{3}+8H_{2}H_{3}\gamma\dot{H}_{3}+8H_{1}H_{3}\gamma\dot{H}_{3}+8H_{2}^{2}\gamma\dot{H}_{3}+16H_{1}H_{2}\gamma\dot{H}_{3}+8H_{1}^{2}\gamma\dot{H}_{3}-4\beta H_{3}^{2}\dot{H}_{3}\\&+10 H_2\beta H_{3}\dot{H}_{3}+10H_{1}\beta H_{3}\dot{H}_{3}+32\alpha H_{2}H_{3}\dot{H}_{3}+32H_{1}\alpha H_{3}\dot{H}_{3}+2H_{2}^{2}\beta\dot{H}_{3}+8H_{1}H_{2}\beta\dot{H}_{3}\\&+16 H_1\alpha H_{2}\dot{H}_{3}+20\gamma\dot{H}_{2}^{2}+6\beta\dot{H}_{2}^{2}+4\alpha \dot{H}_{2}^{2}+24\gamma\dot{H}_{1}\dot{H}_{2}+6\beta\dot{H}_{1}\dot{H}_{2}+16\gamma h_{1}^{2}\dot{H}_{2}+4\beta h_{1}^{2}\dot{H}_{2}\\&+8H_{3}\gamma h_{1}\dot{H}_{2}+32H_{2}\gamma h_{1}\dot{H}_{2}+24H_{1}\gamma h_{1}\dot{H}_{2}+2\beta H_{3}h_{1}\dot{H}_{2}+8H_{2}\beta h_{1}\dot{H}_{2}+4H_{1}\beta h_{1}\dot{H}_{2}\\&+8H_{1}H_{3}\gamma\dot{H}_{2}+24H_{2}^{2}\gamma\dot{H}_{2}+32H_{1}H_{2}\gamma \dot{H}_2+16H_{1}^{2}\gamma\dot{H}_{2}+2\beta H_{3}^{2}\dot{H}_{2}+8\alpha  H_3^{2}\dot{H}_{2}-2H_{2}\beta H_{3}\dot{H}_{2}\\&-16\alpha H_{2}H_{3}\dot{H}_{2}+8H_{2}^{2}\beta\dot{H}_{2}+8 H_1H_{2}\beta\dot{H}_{2}+4H_{1}^{2}\beta\dot{H}_{2}+8\alpha H_{2}^{2}\dot{H}_{2}+20\gamma\dot{H}_{1}^{2}+6\beta\dot{H}_{1}^{2}+4\alpha\dot{H}_{1}^{2}\\&+4\beta h_{1}^{2}\dot{H}_{1}+8H_{3}\gamma h_{1}\dot{H}_{1}+24H_{2}\gamma h_{1}\dot{H}_{1}+32H_{1}\gamma h_{1}\dot{H}_{1}+2\beta H_{3}h_{1}\dot{H}_{1}+4H_{2}\beta h_{1}\dot{H}_{1}+8H_{1}\beta h_{1}\dot{H}_{1}\\&+8H_{2}H_{3}\gamma\dot{H}_{1}+8H_{1}H_{3}\gamma\dot{H}_{1}+16H_{2}^{2}\gamma\dot{H}_{1}+32H_{1}H_{2}\gamma \dot{H}_1+24H_{1}^{2}\gamma\dot{H}_{1}+2\beta H_{3}^{2}\dot{H}_{1}+8\alpha  H_3^{2}\dot{H}_{1}\\&-2H_{1}\beta H_{3}\dot{H}_{1}-16H_{1}\alpha H_{3}\dot{H}_{1}+4H_{2}^{2}\beta\dot{H}_{1}+8 H_1H_{2}\beta\dot{H}_{1}+8H_{1}^{2}\beta\dot{H}_{1}+8H_{1}^{2}\alpha\dot{H}_{1}+4\gamma h_{1}^{4}+2\beta h_{1}^{4}\\&+8H_{2}\gamma h_{1}^{3}+8H_{1}\gamma h_{1}^{3}-2\beta H_{3}h_{1}^{3}-8\alpha H_{3}h_{1}^{3}+2H_{2}\beta h_{1}^{3}+2H_{1}\beta h_1^{3}-4H_{3}^{2}\gamma h_{1}^{2}+12H_{2}^{2}\gamma h_{1}^{2}\\&+12H_{1}^{2}\gamma h_{1}^{2}+4\alpha  H_3^{2}h_{1}^{2}-2H_{2}\beta H_{3}h_{1}^{2}-2H_{1}\beta  H_3h_{1}^{2}-8\alpha H_{2}H_{3}h_{1}^{2}-8H_{1}\alpha H_{3}h_{1}^{2}+4H_{2}^{2}\beta\\&+4\alpha H_{2}^{2}h_{1}^{2}+4H_{1}^{2}\alpha h_{1}^{2}-8H_{3}^{3}\gamma h_{1}-8H_{2}H_{3}^{2}\gamma h_{1}-8H_{1}H_{3}^{2}\gamma h_{1}+8H_{2}^{3}\gamma h_{1}+16H_{1}H_{2}^{2}\gamma h_{1}\\&+8H_{1}^{3}\gamma h_{1}-2\beta H_{3}^{3}h_{1}+2H_{2}\beta H_{3}^{2}h_1+2H_{1}\beta H_{3}^{2}h_{1}+16\alpha H_{2}H_{3}^{2}h_1+16H_{1}\alpha H_{3}^{2}h_{1}-2H_{2}^{2}\beta H_{3}h_{1}\\&-8\alpha H_{2}^{2}H_{3}h_{1}-8 H_1^{2}\alpha H_{3}h_{1}+2H_{2}^{3}\beta h_{1}+2H_{1}H_{2}^{2}\beta h_{1}+2H_{1}^{2}H_{2}\beta h_{1}+2H_{1}^{3}\beta h_{1}-4H_{3}^{4}\gamma-8H_{2}H_{3}^{3}\gamma\\&-4H_{2}^{2}H_{3}^{2}\gamma-8H_{1}H_{2}H_{3}^{2}\gamma-4H_{1}^{2}H_{3}^{2}\gamma+4H_{2}^{4}\gamma+8H_{1}H_{2}^{3}\gamma+12H_{1}^{2}H_{2}^{2}\gamma+8H_{1}^{3}H_{2}\gamma+4H_{1}^{4}\gamma -2\beta H_{3}^{4}\\&-2H_{2}\beta H_{3}^{3}-2 H_1\beta H_{3}^{3}+2H_{1}H_{2}\beta H_{3}^{2}+4\alpha H_{2}^{2}H_{3}^{2}+16H_{1}\alpha H_{2}H_{3}^{2}+4H_{1}^{2}\alpha  H_3^{2}-2H_{2}^{3}\beta H_{3}\\&-2H_{1}^{3}\beta H_{3}-8\alpha H_{2}^{3}H_{3}-8H_{1}\alpha H_{2}^{2}H_{3}-8H_{1}^{2}\alpha H_{2} H_3-8H_{1}^{3}\alpha H_{3}+2H_{2}^{4}\beta+2H_{1}H_{2}^{3}\beta +4H_{1}^{2}H_{2}^{2}\beta\\&+2H_{1}^{4}\beta+4\alpha H_{2}^{4}+4H_{1}^{2}\alpha H_{2}^{2}+4H_{1}^{4}\alpha +2H_{1}\beta H_{3}\dot{H}_{2}+16\gamma h_{1}^{2}\dot{H}_{1}+2H_{2}\beta H_{3}\dot{H}_{1}-2 H_1^{2}H_{2}\beta H_{3}\\&+16H_{1}\alpha h_{1}\dot{H}_{3}-8\alpha H_{3}^{2}\dot{H}_{3}-2H_{1}^{2}\beta H_{3}h_{1}-8H_{1}H_{3}^{3}\gamma-4\alpha H_{3}^{4}+2H_{1}^{3}H_{2}\beta+4H_{1}^{2}\beta h_{1}^{2}+8H_{2}H_{3}\gamma\dot{H}_{2}\\&+2H_{1}^{2}\beta\dot{H}_{3}+16 H_1H_{2}\gamma h_{1}^{2}h_{1}^{2}+2H_{1}H_{2}\beta h_{1}^{2}+16H_{1}^{2}H_{2}\gamma h_{1}-2H_{1}H_{2}^{2}\beta H_{3}
\end{align*}
\begin{align*}
    &\frac{E_{4}}{\sqrt{-g}}=(8\gamma+2\beta)\dddot{H}_{1}+(8\gamma+2\beta)\dddot{H}_{2}+(8\gamma+2\beta)\dddot{H}_{3}+(8\gamma+4\beta+8\alpha)\dddot{h}_{1}+\frac{\Lambda}{8\pi G}+\frac{\dot{H}_{3}}{8\pi G}\\&+\frac{\dot{H}_{2}}{8\pi G}+\frac{\dot{H}_{1}}{8\pi G}+\frac{H_{3}^{2}}{8\pi G}+\frac{H_{2}H_{3}}{8\pi G}+\frac{H_{1}H_{3}}{8\pi G}+\frac{H_{2}^{2}}{8\pi G}+\frac{H_{1}H_{2}}{8\pi G}+\frac{H_{1}^{2}}{8\pi G}+16\gamma h_{1}\ddot{h}_{1}+8\beta h_{1}\ddot{h}_{1}\\&+16H_{2}\gamma\ddot{h}_{1}+16H_{1}\gamma\ddot{h}_{1}+8\beta H_{3}\ddot{h}_{1}+16\alpha H_{3}\ddot{h}_{1}+8H_{2}\beta\ddot{h}_{1}+8H_{1}\beta\ddot{h}_{1}+16\alpha H_{2}\ddot{h}_1+16H_{1}\alpha\ddot{h}_{1}\\&+8\gamma h_{1}\ddot{H}_{3}+2\beta h_{1}\ddot{H}_{3}+24H_{3}\gamma\ddot{H}_{3}+16H_{2}\gamma\ddot{H}_{3}+16H_{1}\gamma\ddot{H}_{3}+6\beta H_{3}\ddot{H}_{3}+4H_{2}\beta\ddot{H}_{3}+4 H_1\beta\ddot{H}_{3}\\&+2\beta h_{1}\ddot{H}_{2}+16H_{3}\gamma\ddot{H}_{2}+24H_{2}\gamma\ddot{H}_{2}+16H_{1}\gamma\ddot{H}_{2}+4\beta H_{3}\ddot{H}_{2}+6H_{2}\beta\ddot{H}_{2}+4H_{1}\beta\ddot{H}_{2}+8\gamma h_{1}\ddot{H}_{1}\\&+16H_{3}\gamma \ddot{H}_1+16H_{2}\gamma\ddot{H}_{1}+24H_{1}\gamma\ddot{H}_{1}+4\beta H_{3}\ddot{H}_{1}+4H_{2}\beta\ddot{H}_{1}+6H_{1}\beta\ddot{H}_{1}+12\gamma\dot{h}_{1}^{2}+6\beta\dot{h}_{1}^{2}+12\alpha\dot{h}_{1}^{2}\\&+16\gamma\dot{H}_{3}\dot{h}_{1}+6\beta \dot{H}_{3}\dot{h}_{1}+8\alpha\dot{H}_{3}\dot{h}_{1}+16\gamma\dot{H}_{2}\dot{h}_{1}+6\beta\dot{H}_{2}\dot{h}_{1}+8\alpha\dot{H}_{2}\dot{h}_{1}+16\gamma\dot{H}_{1}\dot{h}_{1}+6\beta\dot{H}_{1}\dot{h}_{1}\\&-8\gamma h_{1}^{2}\dot{h}_1-4\beta h_{1}^{2}\dot{h}_{1}-8\alpha h_{1}^{2}\dot{h}_{1}+8H_{3}\gamma h_{1}\dot{h}_{1}+8H_{2}\gamma h_{1}\dot{h}_{1}+8H_{1}\gamma h_1\dot{h}_{1}+10\beta H_{3}h_{1}\dot{h}_{1}+32\alpha H_{3}h_{1}\dot{h}_1\\&+10H_{2}\beta h_{1}\dot{h}_{1}+10H_{1}\beta h_{1}\dot{h}_{1}+32\alpha H_{2}h_{1}\dot{h}_{1}+32H_{1}\alpha h_{1}\dot{h}_{1}+8H_{3}^{2}\gamma\dot{h}_{1}+16H_{2}H_{3}\gamma\dot{h}_{1}+16H_{1}H_{3}\gamma \dot{h}_{1}\\&+8H_{2}^{2}\gamma\dot{h}_{1}+16H_{1}H_{2}\gamma\dot{h}_{1}+8H_{1}^{2}\gamma\dot{h}_{1}+2\beta H_{3}^{2}\dot{h}_{1}+8H_{2}\beta  H_3\dot{h}_{1}+8H_{1}\beta H_{3}\dot{h}_{1}+16\alpha H_{2}H_{3}\dot{h}_1\\&+2H_{2}^{2}\beta\dot{h}_{1}+8 H_1H_{2}\beta\dot{h}_{1}+2H_{1}^{2}\beta\dot{h}_{1}+16H_{1}\alpha  H_2\dot{h}_{1}+20\gamma\dot{H}_{3}^{2}+6\beta\dot{H}_{3}^{2}+4\alpha\dot{H}_{3}^{2}+24\gamma\dot{H}_{2}\dot{H}_{3}\\&+24\gamma\dot{H}_{1}\dot{H}_3+6\beta\dot{H}_{1}\dot{H}_{3}+2\beta h_{1}^{2}\dot{H}_{3}+8\alpha h_{1}^{2}\dot{H}_{3}+8H_{3}\gamma h_{1}\dot{H}_{3}+8H_{2}\gamma h_{1}\dot{H}_{3}+8H_{1}\gamma h_{1}\dot{H}_{3}\\&-16\alpha H_{3}h_{1}\dot{H}_{3}+2H_{2}\beta h_{1}\dot{H}_{3}+2H_{1}\beta h_{1}\dot{H}_{3}+24H_{3}^{2}\gamma\dot{H}_{3}+32H_{2}H_{3}\gamma \dot{H}_3+32H_{1}H_{3}\gamma\dot{H}_{3}\\&+24 H_1H_{2}\gamma\dot{H}_{3}+16H_{1}^{2}\gamma\dot{H}_{3}+8\beta H_{3}^{2}\dot{H}_{3}+8\alpha H_{3}^{2}\dot{H}_{3}+8H_{2}\beta H_{3}\dot{H}_{3}+8 H_1\beta H_{3}\dot{H}_{3}+4H_{2}^{2}\beta\dot{H}_{3}\\&+4H_{1}^{2}\beta\dot{H}_{3}+20\gamma\dot{H}_{2}^{2}+6\beta \dot{H}_2^{2}+4\alpha\dot{H}_{2}^{2}+24\gamma\dot{H}_{1}\dot{H}_{2}+6\beta\dot{H}_{1}\dot{H}_2+2\beta h_{1}^{2}\dot{H}_{2}+8\alpha h_{1}^{2}\dot{H}_{2}+8H_{3}\gamma h_{1}\dot{H}_{2}\\&+8H_{2}\gamma h_{1}\dot{H}_{2}+8H_{1}\gamma h_1\dot{H}_{2}+2\beta H_{3}h_{1}\dot{H}_{2}-2H_{2}\beta h_{1}\dot{H}_{2}+2H_{1}\beta h_{1}\dot{H}_{2}-16\alpha H_{2}h_{1}\dot{H}_{2}+16 H_3^{2}\gamma\dot{H}_{2}\\&+32H_{2}H_{3}\gamma\dot{H}_{2}+24H_{1}H_{3}\gamma\dot{H}_{2}+24H_{2}^{2}\gamma\dot{H}_{2}+32H_{1}H_{2}\gamma \dot{H}_2+16H_{1}^{2}\gamma\dot{H}_{2}+4\beta H_{3}^{2}\dot{H}_{2}+8H_{2}\beta H_{3}\dot{H}_{2}\\&+4H_{1}\beta H_{3}\dot{H}_{2}+8H_{2}^{2}\beta\dot{H}_{2}+8H_{1}H_{2}\beta\dot{H}_{2}+4H_{1}^{2}\beta\dot{H}_{2}+8\alpha H_{2}^{2}\dot{H}_{2}+20\gamma\dot{H}_{1}^{2}+6\beta\dot{H}_{1}^{2}+4\alpha \dot{H}_1^{2}\\&+8\alpha h_{1}^{2}\dot{H}_{1}+8H_{3}\gamma h_{1}\dot{H}_{1}+8H_{2}\gamma h_{1}\dot{H}_{1}+8H_{1}\gamma h_1\dot{H}_{1}+2\beta H_{3}h_{1}\dot{H}_{1}+2H_{2}\beta h_{1}\dot{H}_{1}-2H_{1}\beta h_{1}\dot{H}_{1}\\&-16H_{1}\alpha h_{1}\dot{H}_{1}+16 H_3^{2}\gamma\dot{H}_{1}+24H_{2}H_{3}\gamma\dot{H}_{1}+32H_{1}H_{3}\gamma\dot{H}_{1}+16H_{2}^{2}\gamma\dot{H}_{1}+32H_{1}H_{2}\gamma \dot{H}_1+24H_{1}^{2}\gamma\dot{H}_{1}\\&+4\beta H_{3}^{2}\dot{H}_{1}+4H_{2}\beta H_{3}\dot{H}_{1}+8H_{1}\beta H_{3}\dot{H}_{1}+4H_{2}^{2}\beta\dot{H}_{1}+8H_{1}H_{2}\beta\dot{H}_{1}+8H_{1}^{2}\beta\dot{H}_{1}+8 H_1^{2}\alpha\dot{H}_{1}-4\gamma h_{1}^{4}\\&-2\beta h_{1}^{4}-4\alpha h_1^{4}-8H_{3}\gamma h_{1}^{3}-8H_{2}\gamma h_{1}^{3}-8H_{1}\gamma h_{1}^{3}-2\beta H_{3}h_{1}^{3}-2H_{2}\beta h_{1}^{3}-2 H_1\beta h_{1}^{3}-4H_{3}^{2}\gamma h_{1}^{2}\\&-8H_{1}H_{3}\gamma h_{1}^{2}-4H_{2}^{2}\gamma h_{1}^{2}-8H_{1}H_{2}\gamma h_{1}^{2}-4H_{1}^{2}\gamma h_{1}^{2}+4\alpha H_{3}^{2}h_{1}^{2}+2H_{2}\beta H_{3}h_{1}^{2}+2H_{1}\beta H_{3}h_{1}^{2}\\&+16H_{1}\alpha  H_3h_{1}^{2}+2H_{1}H_{2}\beta h_{1}^{2}+4\alpha H_{2}^{2}h_1^{2}+16H_{1}\alpha H_{2}h_{1}^{2}+4H_{1}^{2}\alpha h_{1}^{2}-2\beta H_{3}^{3}h_{1}-8\alpha H_{3}^{3}h_{1}-2H_{2}\beta  H_3^{2}h_{1}\\&-2H_{1}\beta H_{3}^{2}h_{1}-8\alpha H_{2}H_{3}^{2}h_{1}-8H_{1}\alpha H_{3}^{2}h_{1}-2H_{2}^{2}\beta H_{3}h_1-2H_{1}^{2}\beta H_{3}h_{1}-8\alpha H_{2}^{2}H_{3}h_{1}-8H_{1}^{2}\alpha H_{3}h_{1}\\&-2H_{2}^{3}\beta h_{1}-2H_{1}H_{2}^{2}\beta h_{1}-2H_{1}^{2}H_{2}\beta h_{1}-2H_{1}^{3}\beta h_{1}-8\alpha H_{2}^{3}h_{1}-8H_{1}\alpha H_{2}^{2}h_{1}-8H_{1}^{2}\alpha H_{2}h_{1}-8H_{1}^{3}\alpha h_{1}\\&+4H_{3}^{4}\gamma+8H_{2}H_{3}^{3}\gamma+8H_{1}H_{3}^{3}\gamma+12H_{2}^{2}H_{3}^{2}\gamma+16H_{1}H_{2}H_{3}^{2}\gamma+12H_{1}^{2} H_3^{2}\gamma+8H_{2}^{3}H_{3}\gamma+16H_{1}H_{2}^{2}H_{3}\gamma\\&+16H_{1}^{2}H_{2}H_{3}\gamma+8H_{1}^{3}H_{3}\gamma+4H_{2}^{4}\gamma+8H_{1}H_{2}^{3}\gamma+12H_{1}^{2}H_{2}^{2}\gamma+8 H_1^{3}H_{2}\gamma+4H_{1}^{4}\gamma+2\beta H_{3}^{4}+4\alpha  H_3^{4}\\&+2H_{1}\beta H_{3}^{3}+4H_{2}^{2}\beta H_{3}^{2}+2H_{1}H_{2}\beta H_{3}^{2}+4H_{1}^{2}\beta  H_3^{2}+4\alpha H_{2}^{2}H_{3}^{2}+4H_{1}^{2}\alpha H_{3}^{2}+2H_{2}^{3}\beta H_{3}+2H_{1}H_{2}^{2}\beta H_{3}\\&+2H_{1}^{2}H_{2}\beta H_{3}+2H_{1}^{3}\beta H_{3}+2H_{2}^{4}\beta+2H_{1} H_2^{3}\beta+4H_{1}^{2}H_{2}^{2}\beta+2H_{1}^{3}H_{2}\beta+2 H_1^{4}\beta+4\alpha H_{2}^{4}+4H_{1}^{2}\alpha H_{2}^{2}\\&+16\alpha h_{1}\ddot{h}_{1}+16H_{3}\gamma\ddot{h}_{1}+8\gamma h_{1}\ddot{H}_{2}+2\beta h_{1}\ddot{H}_{1}+8\alpha\dot{H}_{1}\dot{h}_{1}+16H_{1}\alpha H_{3}\dot{h}_{1}+4H_{1}H_{2}\beta\dot{H}_{3}+2H_{2}\beta H_{3}^{3}\\&+6\beta\dot{H}_{2}\dot{H}_{3}-2\beta H_{3}h_{1}\dot{H}_{3}+16H_{2}^{2}\gamma\dot{H}_{3}+2\beta h_{1}^{2}\dot{H}_{1}-8H_{2}H_{3}\gamma h_{1}^{2}+16\alpha H_{2}H_{3}h_{1}^{2}+4H_{1}^{4}\alpha
\end{align*}

\end{document}